\documentclass[prl,showpacs,twocolumn,floats,10pt,aps,citeautoscript,longbibliography,superscriptaddress]{revtex4-2}

\usepackage[T1]{fontenc}
\usepackage[utf8]{inputenc}
\usepackage{amsmath,amssymb,bm}
\usepackage{graphicx}
\usepackage{xcolor}

\RequirePackage[
  hyperindex,colorlinks,bookmarksnumbered,
  plainpages=true,pdfstartview=FitH]{hyperref}
\hypersetup{linkcolor=blue,urlcolor=blue,citecolor=blue} 
\usepackage[all]{hypcap}

\usepackage{placeins}
\usepackage{lipsum}
\usepackage{xcolor}
\usepackage[normalem]{ulem}
\usepackage{comment}
\usepackage{graphicx}
\usepackage{wrapfig}
\usepackage{dcolumn}
\usepackage{bm}

\usepackage{blindtext}
\usepackage{graphics}
\usepackage{verbatim}   
\usepackage[ruled,lined]{algorithm2e}
\usepackage{amsfonts}
\usepackage{amsmath}
\usepackage{amssymb}
\usepackage{mathrsfs} 
\usepackage{multirow}
\usepackage{physics}
\usepackage{bbm}  
\usepackage{adjustbox}
\usepackage{appendix}

\usepackage{tabstackengine}
\setstackEOL{\cr}

\newcommand{\Eq}[1]{Eq.~\eqref{#1}}

\newcommand{\Fig}[1]{Fig.~\ref{#1}}
\newcommand{\Figs}[1]{Figs.~\ref{#1}}

\newcommand{\mr}[1]{\mathrm{#1}}

\newcommand{\mc}[1]{\mathcal{#1}}

\newcommand{\bK}{{\boldsymbol{\mathrm{K}}}}
\newcommand{\bk}{{\boldsymbol{\mathrm{k}}}}
\newcommand{\br}{{\boldsymbol{\mathrm{r}}}}
\newcommand{\bri}{{\boldsymbol{\mathrm{r}}}}

\newcommand{\pdag}{\phantom{\dag}}

\newcommand{\FermiL}{FL}
\newcommand{\TNFL}{T_{\mr{NFL}}}
\newcommand{\TFL}{T_{\mr{FL}}}

\newcommand{\ii}{\mr{i}}

\newcommand{\bubble}{{\mr{B}}} 
\newcommand{\vertex}{{\mr{V}}} 

\newcommand{\LSCO}{$\mr{La}_{2-x}\mr{Sr}_x\mr{CuO}_4$}

\newcommand{\LCO}{$\mr{La}_2\mr{CuO}_4$}

\definecolor{darkgreen}{rgb}{0,0.5,0}
\definecolor{darkblue}{rgb}{0,0,0.4}

\newcommand{\jvd}[1]{{\color{magenta}#1}}

\newcommand{\MP}[1]{{\color{violet}#1}} 

\newcommand{\safeincludegraphics}[2][]{%
  \IfFileExists{#2}{\includegraphics[#1]{#2}}{%
    \fbox{\begin{minipage}[c][0.28\textheight][c]{0.92\linewidth}
    \centering\small Missing figure file:\\[2pt]\texttt{\detokenize{#2}}
    \end{minipage}}}}

\newcommand{\LMUMunich}{Arnold Sommerfeld Center for Theoretical Physics, Center for NanoScience, and Munich Center for Quantum Science and Technology, Ludwig-Maximilians-Universit\"at M\"unchen, 80333 Munich, Germany}
\newcommand{\RutgersUniversity}{Department of Physics and Astronomy and Center for Condensed Matter Theory, Rutgers University, Piscataway, New Jersey 08854-8019, USA}

\makeatletter
\def\maketitle{
\@author@finish
\title@column\titleblock@produce
\suppressfloats[t]}
\makeatother

\begin{document}

\title{Dynamical scaling near the pseudogap quantum critical point of the two-dimensional Hubbard model}

\author{Mathias Pelz}
\email{mathias.pelz@lmu.de}
\affiliation{\LMUMunich}

\author{Gabriel Kotliar}
\email{kotliar@physics.rutgers.edu}
\affiliation{\RutgersUniversity}

\author{Jan von Delft}
\email{vondelft@lmu.de}
\affiliation{\LMUMunich}

\author{Andreas Gleis}
\email{andreas.gleis@rutgers.edu}
\affiliation{\RutgersUniversity}

\date{\today}

\begin{abstract}
We study dynamical scaling in the quantum-critical fan of the pseudogap-metal to Fermi-liquid transition of the two-dimensional Hubbard model. 
Using a four-patch dynamical cluster approximation with the numerical renormalization group as a cluster impurity solver, we access real-frequency dynamics over several decades at arbitrary temperatures. 
Close to the critical doping,
the local spin and cluster-current susceptibility spectra exhibit $x=\omega/T$ scaling of the form $\chi''(\omega,T)\sim \tanh(x/2)$, and the cluster contribution to the optical conductivity obeys $T\sigma'_{\rm cl}(\omega,T) \sim \tanh(x/2)/x$, implying a $1/T$ cluster dc conductivity.
In the scaling regime, the vertex contribution to the cluster optical response is much larger than the bubble contribution.
We further find evidence for a marginal-Fermi-liquid nodal self-energy. This,  together with the $1/T$ vertex contribution to the conductivity,
implies
strange-metal optical transport in the quantum critical region.
Our results describe several qualitative aspects of several experimental observations. 
\end{abstract}

\maketitle

\textit{Introduction.---}
Hole-doped cuprates remain the superconductors with the highest transition temperatures at ambient pressure. 
Yet, almost four decades after their discovery, central aspects of their phase diagram, such as the underdoped pseudogap~(PG) metal and the strange-metal regime near optimal doping,  still lack a universally agreed-upon microscopic explanation \cite{Keimer2015}.

It was suggested early on that the superconducting dome conceals a quantum critical point~(QCP) at a critical doping $p^{\ast}$, separating the underdoped PG metal from the overdoped Fermi liquid (\FermiL)~\cite{Castellani1995,Varma1997,Varma1999,Tallon1999,Sachdev2010}. 
This conjecture is supported experimentally by a singular specific heat coefficient~\cite{Michon2019} and a sharp transformation of the Fermi surface, indicated by Hall-effect~\cite{Badoux2016,Collignon2017_PRB}, angle-dependent magnetoresistance~\cite{Fang2022_ADMR}, quantum oscillation~\cite{DoironLeyraud2007,Sebastian2012}, and ARPES measurements~\cite{Reber2019}. 
The latter reveals that low-energy electronic spectral weight is lost over a large part of the antinodal Fermi surface when decreasing doping 
from the overdoped \FermiL\ into the underdoped PG metal, while the nodal region remains comparatively coherent~\cite{Norman1998,Damascelli2003,He2011,Vishik2012,Chen2019,Reber2019}. 
Further, from this perspective, the strange metal observed near optimal doping is interpreted as the quantum-critical fan emanating from this QCP~\cite{Valla1999}.
Experimental signatures of the strange metal include linear-in-$T$ resistivity at low temperatures~\cite{Cooper2009,Legros2019,Grissonnanche2021}, a logarithmic temperature dependence of the specific-heat coefficient~\cite{Michon2019}, and dynamical $\omega/T$ scaling of magnetic~\cite{Keimer1991,Hayden1991,Aeppli1997,Radaelli2026}, charge~\cite{Guo2024_Conformal}, and optical responses~\cite{VanderMarel2003,VanderMarel2006,Michon2023}.
The extended quantum critical region~\cite{Reber2019,Ayres2021} found in some experiments has been interpreted to be the result of disorder~\cite{Patel2024}.

However, the aforementioned phenomena are difficult to reconcile with a Hertz--Millis--Moriya theory~\cite{Hertz1976,Millis1993,Moriya1985} without additional ingredients, and the nature of the putative QCP remains unclear.
It is still debated whether the transition is driven by a conventional (fluctuating) order parameter~\cite{Emery1995,Eberlein2016,Verret2017,Bonetti2020,Bonetti2022,Klett2022,Lihm2026,Forni2026}, by fractionalization or topological order~\cite{Senthil2003,Senthil2004,Moon2011,Scheurer2018,Wu2018,Zhang2020,Mascot2022,Wang2022,Wu2024,Bonetti2026}, or whether disorder plays an important role~\cite{Grilli2022,Caprara2022,Patel2023,Li2024_YSYK,PatelLuntsSachdev2024}.
An early phenomenological approach to strange metal dynamics was provided by the marginal Fermi liquid~(MFL) proposal of Varma \textit{et al.}~\cite{Varma1989_MFL}.
There, bosonic fluctuations across many momenta exhibiting spectra proportional to $\tanh(\omega/2T)$ produce a single-particle scattering rate proportional to $\max(|\omega|,T)$ and give rise to a logarithmic temperature dependence of the specific-heat coefficient. 
Further, if vertex contributions to the conductivity are irrelevant, this gives rise to a linear-in-$T$ resistivity.
Recently, the Yukawa--Sachdev--Ye--Kitaev~(YSYK) construction~\cite{Patel2023,Li2024_YSYK,PatelLuntsSachdev2024} has been put forward as a universal mechanism towards MFL physics, though spatially disordered interactions are required to relax momentum and ensure irrelevant vertex contributions.
In clean lattices, current relaxation requires a separate mechanism, for instance, through Umklapp scattering~\cite{HartnollHofman2012,Rice2017_Umklapp,Lee2021_Umklapp}.

In this Letter, we study the quantum-critical region of the PG-metal to Fermi-liquid QCP in the one-band Hubbard model, using a four-patch dynamical cluster approximation~(DCA), a cluster extension of dynamical mean-field theory~(DMFT)~\cite{Georges1996_DMFT,Hettler1998_DCA,Hettler2000_DCA,Maier2005_clusters}. 
It has been established in a large body of previous work that cluster DMFT approximations capture central aspects of the hole-doped cuprates, in particular the existence of a PG metal~\cite{Huscroft2001,Civelli2005_FermiSurfaceBreakup,Maier2005_clusters,Tremblay2006_num,Kyung2006,Stanescu2006_FermiArcs,Macridin2006,Haule2007,Ferrero2009_FermiArcs,Werner2009,Liebsch2009,Mikelsons2009,Vidhyadhiraja2009,Gull2010_patching-pg,Khatami2010,Yang2011,Sordi2010,Sordi2011,Sordi2012_Widom,Sordi2012_PRL,Sordi2013,Sordi2019,Scheurer2018,Wu2018,Meixner2024}.
Further, it seems that a four-site approach is enough to capture the qualitative aspects~\cite{Hettler2000_DCA,Civelli2005_FermiSurfaceBreakup,Stanescu2006_FermiArcs,Gull2010_patching-pg,Sordi2010,Sordi2011,Sordi2012_Widom,Sordi2012_PRL,Sordi2013,Sordi2019}.
However, the existence of a QCP separating PG metal from \FermiL\ has been controversial, with some work providing evidence for a QCP~\cite{Vidhyadhiraja2009,Liebsch2009,Mikelsons2009,Khatami2010,Yang2011}.
By contrast, Sordi~\textit{et al.}~\cite{Sordi2010,Sordi2011,Sordi2012_Widom,Sordi2012_PRL,Sordi2013,Sordi2019} have established in a certain parameter region ($t' = 0$, $U/t \leq 6$) that the transition is first-order, and criticality is associated with a finite (but low) temperature critical endpoint.

In Ref.~\cite{Pelz2026_QCP}, we have made significant progress on this matter, by using four-patch DCA in conjunction with the numerical renormalization group~(NRG)~\cite{Wilson1975_NRG,Bulla2008_NRG,Anders2005_NRG,Lee2016_NRG,Kugler2022_SEtrick,Weichselbaum2012b_QSpace}.
NRG is well-suited for this purpose, since it offers unprecedented real-frequency spectral resolution at arbitrarily low energies and temperatures. In a parameter regime relevant for cuprates, this has allowed us to establish the existence of a QCP that occurs between two distinct {\FermiL}s (one is a PG metal) with different Fermi surfaces and quasiparticle content.
At this transition, the Fermi-liquid scale is continuously suppressed to zero, suggesting that the QCP is governed by a non-Fermi-liquid (NFL) fixed point. Here, we study dynamical scaling in the quantum-critical fan emanating from this NFL fixed point.

\textit{Model and methods.---}
In this work, we aim to study universal dynamical phenomena in the quantum critical region of the pseudogap QCP in the cuprates.
For that, we consider the two-dimensional square-lattice $t$-$t'$-$U$ Hubbard model, which is believed to contain the necessary ingredients to describe cuprate physics~\cite{Zhang1988},
\begin{align}
    H=&-\sum_{\br \br' \sigma} t_{\br \br'} c_{\br\sigma}^{\dagger}c_{\br'\sigma}^{\phantom{\dagger}}
    +U\sum_{\br} n_{\br \uparrow}n_{\br\downarrow}
    -\mu\sum_{\br \sigma} n_{\br \sigma} \, .
\label{eq:Hubbard}
\end{align}%
The hopping matrix $t_{\br \br'}$ contains nearest and next-nearest neighbor matrix elements, denoted $t$ and $t'$, respectively.
We use $t' = -0.3 t$ and $U = 7 t$, appropriate for describing cuprates~\cite{Hirayama2018,Schmid2023}.
In the following, we set $t = 1$ as our unit of energy.
By tuning the chemical potential $\mu$, we change the hole doping $p$ (filling $n = 1-p$) to drive this model through a quantum phase transition.

Physical quantities are computed within a four-patch DCA approximation, using the star patching scheme of Refs.~\cite{Gull2010_patching-pg,Pelz2026_QCP}.
DCA approximates the self-energy as momentum-independent within each patch, which leads to discontinuities at the patch boundaries.
The four patches are centered around momenta $\bK\in\{(0,0),(\pi,0),(0,\pi),(\pi,\pi)\}$, which are also used as patch labels.
The self-energy is computed from an effective four-impurity model, which we solve using NRG.
Using the QSpace-based~\cite{Weichselbaum2012a_QSpace,Weichselbaum2020_QSpace,Weichselbaum2024_QSpace} MuNRG package~\cite{Lee2016_NRG,Lee2017,Lee2021}, we exploit the SU(2)$\times$U(1) spin and charge symmetries, use an interleaved Wilson chain with logarithmic discretization parameter $\Lambda = 8$, and keep up to $N_{\mr{keep}} = 3\cdot10^4$ low-energy states during iterative diagonalization, resulting in a rescaled truncation energy of $E_{\mr{trunc}} \geq 5$. 
An interleaved Wilson chain geometry \cite{Mitchell2014_iNRG,Stadler2016_iNRG} is used to fully interleave the bath modes associated with the different patches (labeled by their patch momenta) in the order: $\bK_1 = (0,0)$, $\bK_2 = (0,\pi)$, $\bK_3 = (\pi,0)$, $\bK_4 = (\pi,\pi)$.
We consider only non-symmetry-broken, normal-state solutions.

DCA+NRG yields access to real-frequency electronic spectral functions and self-energies with coarse-grained momentum resolution.
For our analysis, we further compute dynamical susceptibilities of the self-consistent cluster impurity model, which correspond to local or short-ranged dynamical susceptibilities of the lattice model.

Consider the retarded dynamical susceptibility 
of a bosonic cluster operator $O$
(with $\langle \;\; \rangle$ a thermal average):
\begin{align}
    \chi[O](\omega,T) =-\ii\int_0^\infty \dd t\, e^{\ii(\omega+\ii0^+)t}
    \left\langle [O(t),O^{\dagger}]\right\rangle . 
\label{eq:retarded}
\end{align}
We compose it as $\chi = \chi' - \ii \pi \chi''$ into real and spectral parts.
Within NRG, $\chi''$ is computed via the Lehmann representation;
the discrete raw spectrum is broadened with a log-Gaussian kernel~\cite{Lee2016_NRG},
and $\chi'$ is subsequently obtained through a Kramers-Kronig relation.
We further study the corresponding imaginary-time correlator, 
\begin{align}
    \langle O(\tau) O^{\dag} \rangle =\int_{-\infty}^{\infty}\dd\omega\,
    \frac{e^{-\tau\omega}}{1-e^{-\omega/T}}\,\chi''[O](\omega,T) \, .
    \label{eq:tau_transform}
\end{align}
We evaluate it using the discrete raw real-frequency spectrum as input, thus avoiding  broadening artifacts. 

We focus on the local spin $S^z_\bri$ and the current density in $x$-direction on the cluster, $j_{\mr{cl}}^{x}$, defined as 
\begin{subequations}
    \label{eqs:operators}
    \begin{align}
    S^{z}_\bri &= \tfrac{1}{2} (c^{\dag}_{\br\uparrow} c^{\pdag}_{\br\uparrow} -c^{\dag}_{\br \downarrow} c^{\pdag}_{\br \downarrow} ) , 
\\
    j_{\mr{cl}}^{x} &= -\tfrac{1}{2} \mr{i} e 
    \sum_{\br \in R} \sum_{\br' \in L} \sum_{\sigma} 
    t_{\br \br'} (c^{\dag}_{\br,\sigma} c^{\pdag}_{\br',\sigma} - 
    \mr{h.c.}) \, ,
\end{align}
\end{subequations}
where the sums on $\br$ or $\br'$ run over the two right ($R$) or the two left ($L$) sites of the square cluster, respectively. 
\begin{figure}
    \centering
    \includegraphics[width=\linewidth]{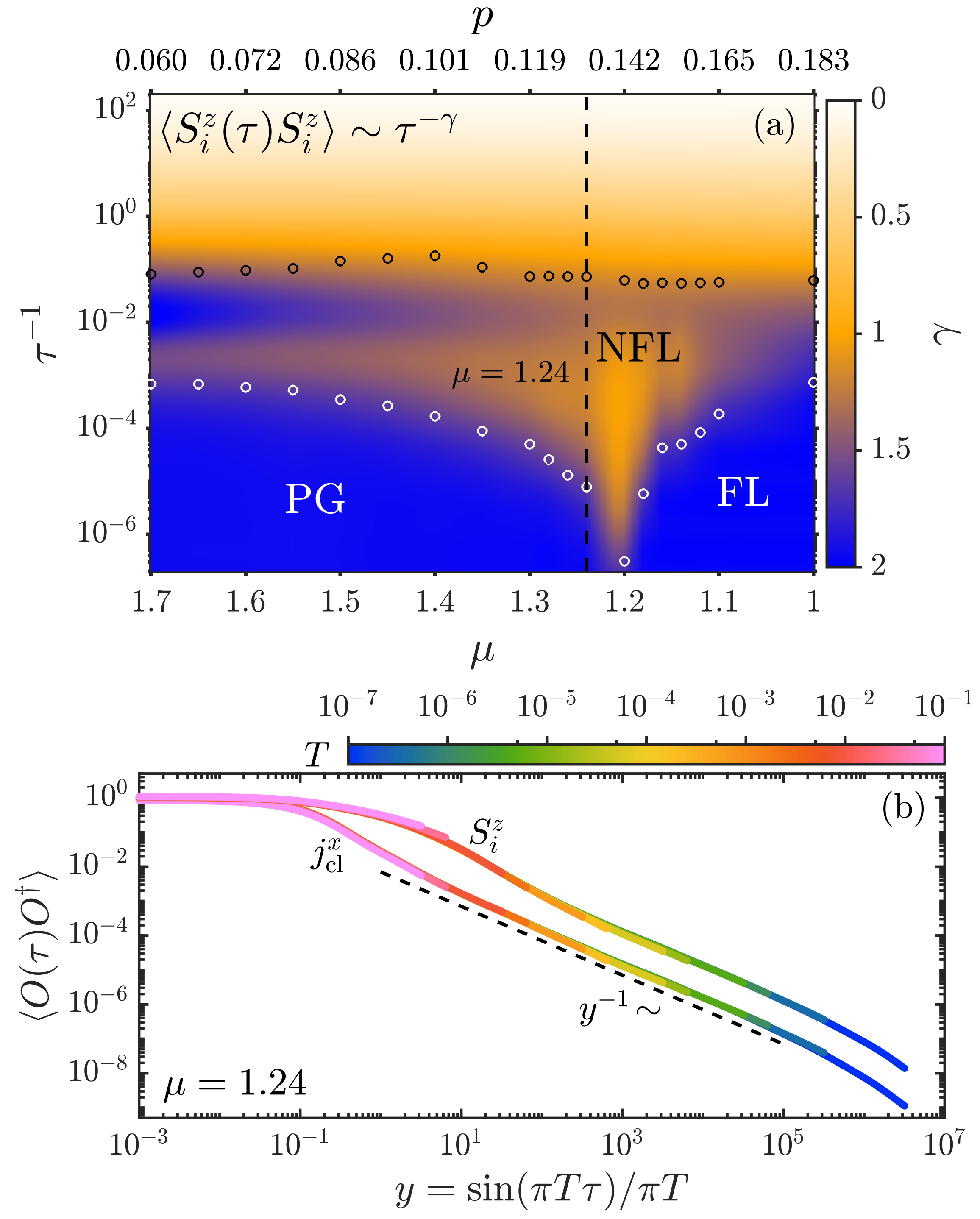}
    \caption{(a) Zero-temperature phase diagram extracted from the power-law decay of the imaginary-time local spin correlator, $\langle S_\bri ^z(\tau)S_\bri ^z\rangle_{T=0}\sim\tau^{-\gamma}$. The color scale shows the exponent $\gamma$ versus chemical potential $\mu$ and inverse imaginary time $\tau^{-1}$. (Top abscissa: the corresponding hole doping $p$ at $T=0$.) Blue regions with $\gamma=2$ indicate Fermi-liquid behavior at long times. The yellow region with $\gamma\simeq1$ identifies the intermediate non-Fermi-liquid region emerging near the critical doping $p^\ast\simeq0.14$. Black and white circles denote the crossover energies $\TNFL$ and $\TFL$, respectively \cite{DefinitionTFL-TNFL}.  The vertical dashed line marks $\mu=1.24$, where we study scaling. (b) Local spin and cluster-current correlators at $\mu=1.24$, plotted vs.\ $y=\sin(\pi T\tau)/\pi T$ for $\tau \in [0,1/2T]$ and 12 temperatures,  indicated by ticks on the color bar. At intermediate temperatures and times, both correlators decay as $y^{-1}$ (dashed line).
    }
    \label{fig:phasedia} \vspace{-5mm}
\end{figure}

\textit{Phase diagram.---}
The zero-temperature phase diagram and energy scales obtained through DCA+NRG are shown in Fig.~\ref{fig:phasedia}(a).
It displays the $T=0$ imaginary-time exponent $\alpha$ of the local spin correlations, $\langle S^{z}_\bri (\tau) S^{z}_\bri  \rangle \sim \tau^{\jvd{-}\MP{\gamma}}$, for different doping levels $p$ and imaginary time scales $\tau$.
At long times, $\tau^{-1} < \TFL$, where $\TFL$ is the \FermiL\ scale (white circles), we find $\MP{\gamma} = 2$ (blue), characteristic of \FermiL\ behavior.
The scale $\TFL$ continuously vanishes at a critical doping $p^{\ast} \simeq 0.14$, where a quantum phase transition between a PG metal with a reconstructed Fermi surface and a pseudogapped electronic spectral function and a normal \FermiL\ occurs~\cite{Pelz2026_QCP}. 

The vanishing \FermiL\ scale in the vicinity of $p^{\ast}$ gives rise to an intermediate NFL region, 
$\TFL < \tau^{-1} < \TNFL$, where $\TNFL$ is the NFL scale (black circles). There, we find an SYK-like exponent, $\gamma= 1$, i.e.\ $\langle S^{z}_\br(\tau) S^{z}_\br \rangle \sim \tau^{-1}$ [yellow in Fig.~\ref{fig:phasedia}(a)].
Our data suggest that this quantum critical region extends to $\tau^{-1} \to 0$ at $p^{\ast}$. This indicates that this region is controlled by an unstable NFL fixed point separating the PG metal from the normal FL.
At short times, $ \tau^{-1} > \TNFL$,  the spin dynamics becomes very slow, with $\gamma < 0.5$, consistent with local-moment behavior.

We henceforth focus on $\mu = 1.24$.
There, the system at $T=0$ is a PG metal with filling $p \simeq 0.135$. The intermediate NFL window extends over $\sim 4$ decades, $[\TFL,\TNFL] = [10^{-5},10^{-1}]$. 
To elucidate the NFL behavior, we study the effects of non-zero temperature on $\langle S^{z}_\bri (\tau) S^{z}_\bri  \rangle$ and $\langle j^{x}_{\mr{cl}} (\tau) j^{x}_{\mr{cl}} \rangle$. 
When $\langle S^{z}_\bri (\tau) S^{z}_\bri  \rangle$ is plotted vs.\ $y = \sin (\pi T \tau)/\pi T$~\cite{Cai2020} [Fig.~\ref{fig:phasedia}(b)],
curves for different temperatures collapse onto a single curve, and likewise for  $\langle j^{x}_{\mr{cl}} (\tau) j^{x}_{\mr{cl}} \rangle$;
and for intermediate temperatures and times,
$T, \tau^{-1} \in [\TFL,\TNFL]$, both correlators show power-law decay, $ \sim y^{-\gamma}$, with $\gamma = 1$.  Thus, both have the scaling form $\sim T^{\gamma} 
\mc{F}(\tau T)$, and in the NFL window, $T$ is the only relevant energy scale, implying Planckian scaling.

\begin{figure}
    \centering
    \includegraphics[width=\linewidth]{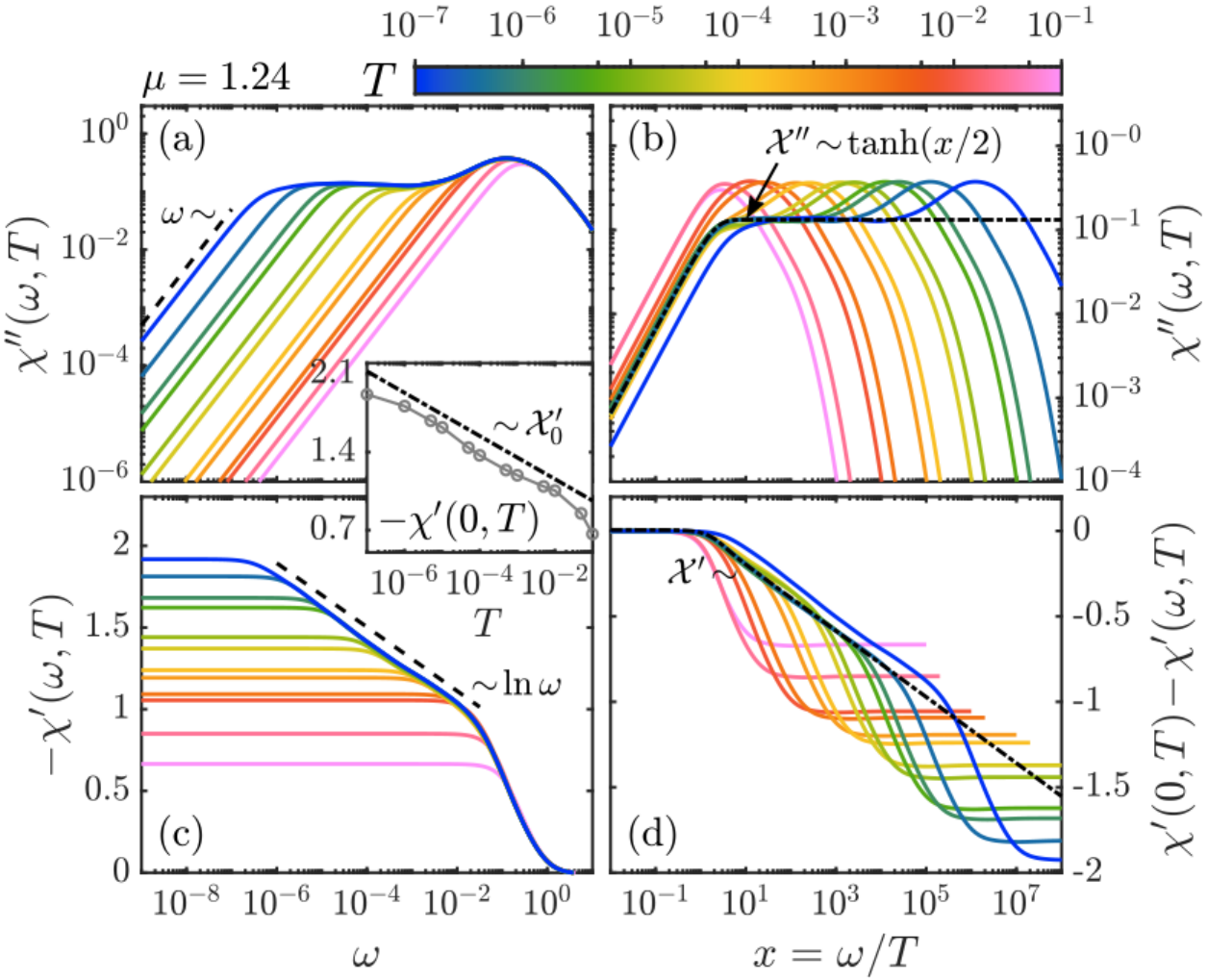}
    \caption{Real-frequency scaling of the local spin susceptibility $\chi[S_\bri ^z](\omega,T)$ at $\mu=1.24$.
    (a) Spectral part $\chi''$ and (c) real part $-\chi'$, plotted vs.\ $\omega$ for 12 temperatures, indicated by ticks on the color bar. Upon lowering the temperature through the NFL window, $\chi''$ develops an extended plateau for $T\lesssim|\omega|\lesssim \TNFL$, while the static response, $\chi'(0,T)$, grows approximately logarithmically (inset).
    (b) The spectral part, plotted as $\chi''(\omega,T)$ vs.\ $x=\omega/T$, shows a scaling collapse onto $\mc{X}''(x)=\mc{X}''_0\tanh(x/2)$. (d) After subtracting the static part, $\chi'$ shows a scaling collapse, too. Black dash-dotted curves depict the scaling functions; dashed lines are guides-to-the-eye.
    }
    \label{fig:Sz_omega} \vspace{-5mm}
\end{figure}

\textit{Dynamical scaling.---}
Such $\tau T$ scaling for imaginary-time correlators implies $\omega/T$ scaling for their real-frequency spectral functions, $\chi''(\omega,T) = T^\jvd{\alpha} \mc{X}''(\omega/T)$, with  $\alpha = \gamma - 1 = 0$. 
For $\mc{F}(\tau T) = \pi/\sin(\pi \tau T)$,   
the scaling function is $\mc{X}''(x) = \mc{X}''_0 \tanh(x/2)$, where $\mc{X}''_0$ is a non-universal prefactor. 
Indeed, when the temperature-dependent spin spectrum, $\chi''[S^{z}_\bri ](\omega,T)$ [Fig.~\ref{fig:Sz_omega}(a)] is plotted vs.\ $x = \omega/T$ [Fig.~\ref{fig:Sz_omega}(b)],  
it exhibits a scaling collapse 
onto $\mc{X}_0^{\prime \prime} 
\tanh(x/2)$ (black dash-dotted line).
This Planckian dynamical scaling sets in around $T \simeq 10^{-2} \text{--} 10^{-3}$ and extends down to $T \simeq 10^{-6}$, implying scaling over 3 to 4 decades.
The corresponding static susceptibility, $\chi'[S^{z}_\bri ](0,T)$ [Fig.~\ref{fig:Sz_omega}, inset] depends on the high-frequency cutoff $\TNFL$ and diverges logarithmically with $T/\TNFL$, $\chi'[S^{z}_\bri ](0,T) \sim \ln T/\TNFL$.
When this static part is subtracted, $\chi'[S^{z}_\bri ](\omega,T) - \chi'[S^{z}_\bri ](0,T)$ [Fig.~\ref{fig:Sz_omega}(c)] exhibits Planckian $\omega/T$ scaling, too [Fig.~\ref{fig:Sz_omega}(d)], with a similar scaling range as $\chi''$ and a scaling function $\mc{X}'(x)$ discussed in  Ref.~\cite{supplement}. 
Our spin susceptibility thus follows the scaling form conjectured by Varma \textit{et al.}~\cite{Varma1989_MFL} via phenomenological considerations.

Experimentally, $\omega/T$ scaling of spin fluctuations in the strange metal region has been measured in doped \LCO~\cite{Keimer1991,Hayden1991,Aeppli1997,Radaelli2026}.
A scaling analysis of the local susceptibility in Ref.~\cite{Keimer1991} revealed an exponent of $\alpha = 0$, consistent with our calculations.
To compare the temperature and frequency scales in the experiments to our data, we assume $t = 0.5 \, \mr{eV}$, 
suitable for most cuprates~\cite{Hirayama2018,Schmid2023}.
Experimentally, the onset for scaling seems to be around $T \simeq 300 \, \mr{K}$ or $\hbar \omega \simeq 15 \, \mr{meV}$, which corresponds to around $10^{-2} \, t$,
matching our scaling onset in order-of-magnitude.

\begin{figure}
    \centering
    \includegraphics[width=\linewidth]{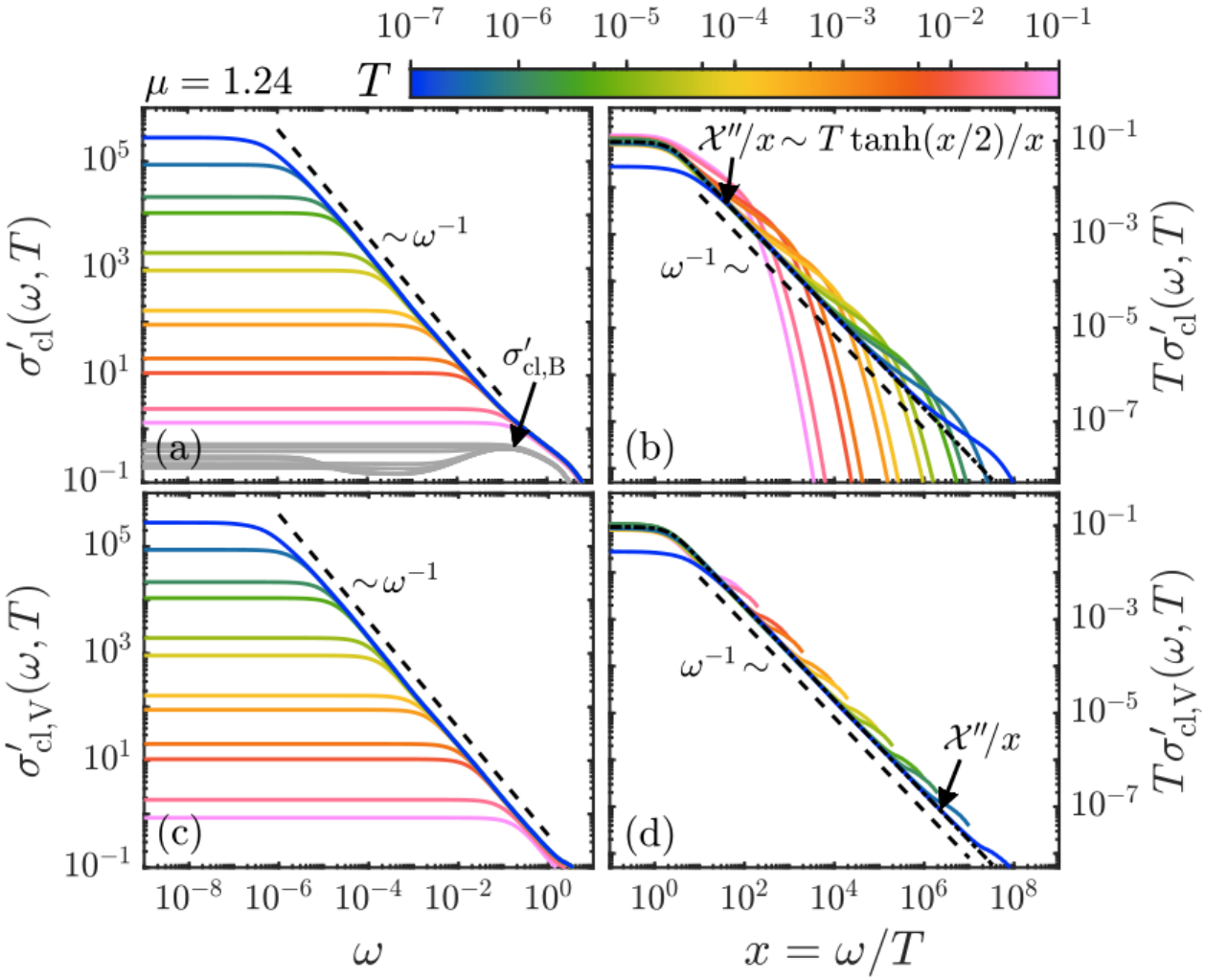}
    \caption{Cluster contribution to the real part of the optical conductivity, $\sigma'_{\rm cl}(\omega,T)=\chi''[j^x_{\rm cl}](\omega,T)/\omega$, at $\mu=1.24$.
    (a) $\sigma'_\mr{cl}(\omega,T)$ vs.\ $\omega$ for 12  temperatures,  indicated by ticks on the color bar;  in the NFL regime, $\sigma'_\mr{cl}$ scales 
    as $\omega^{-1}$ (dashed line).
    Grey curves show the cluster bubble contribution $\sigma'_{\rm{cl},\bubble}$, computed from dressed propagators without vertex contributions.
    (b) $T\sigma'_{\rm cl}(\omega,T)$ vs.\ $x=\omega/T$, showing 
    a scaling collapse onto $\tanh(x/2)/x$ (dash-dotted curve).
    (c) Vertex contribution $\sigma'_{\rm cl,\vertex}=\sigma'_{\rm cl}-\sigma'_{\rm cl,\bubble}$, plotted vs.\ $\omega$.
    In the NFL regime it is orders of magnitude larger than 
    $\sigma'_{\mr{cl},\bubble}$, and (d) obeys the same $\omega/T$ scaling as the full cluster response $\sigma'_{\mr{cl}}$.
    }
    \label{fig:res} \vspace{-5mm}
\end{figure}

\textit{Optical conductivity.---}
One of the most intriguing features of the strange metal observed in cuprates is its remarkably robust 
linear-in-$T$ resistivity and dynamical scaling of the optical conductivity.
The latter is determined by current fluctuations. Since the short-ranged cluster-current correlator shows $\tau T$ scaling similar to the spin correlator [Fig.~\ref{fig:phasedia}(b)], 
the cluster-current spectrum  exhibits $\omega/T$ scaling in the NFL window, too, again of the form $\chi''[j^{x}_{\mr{cl}}](\omega,T) \propto \tanh(\omega/2T)$.
The corresponding cluster contribution to the real part of the optical conductivity, given by 
$\sigma'_{\mr{cl}}(\omega,T) = \chi''[j^{x}_{\mr{cl}}](\omega,T)/\omega$, is shown in Fig.~\ref{fig:res}(a) for different temperatures.
In the NFL window, $T \sigma'_{\mr{cl}}$ plotted vs.\ $x=\omega/T$ indeed exhibits a scaling collapse [Fig.~\ref{fig:res}(b)] onto $\tanh(x/2)/x$ (black dash-dotted line). Remarkably, 
this scaling implies that the cluster contribution to the static conductivity, $\sigma'_{\mr{cl}}(0,T)$, exhibits $1/T$ behavior, reminiscent of the \textit{uniform} conductivity in a strange metal.
The $\omega/T$ scaling of $T\sigma'_\mr{cl}$ sets in at higher temperatures than that of $\chi''[j^x_\br]$ and extends over 5 decades, 
from at least $T = 10^{-1} t$ (highest $T$ considered) down to $T = 10^{-6} t$. Using $t = 0.5 \, \mr{eV}$, this roughly corresponds to temperatures from $600 \, \mr{K}$ down to $6 \, \mr{mK}$.
Experimentally, such a remarkably wide range is observed for the resistivity of \LSCO: it is linear-in-$T$ from around 1000 K~\cite{Gurvitch1987,Ono2007} down to at least 1.5 K~\cite{Cooper2009} if superconductivity is suppressed by a magnetic field. 

To further analyze $\sigma'_{\mr{cl}}(\omega,T)$, we decompose it
as $\sigma'_\mr{cl} = \sigma'_\mr{cl,\bubble} + \sigma'_\mr{cl,\vertex}$
into  bubble and vertex contributions, shown in grey or in color in Figs.~\ref{fig:res}(a)
or~\ref{fig:res}(c), respectively.
Deep in the NFL regime we find $\sigma'_\mr{cl,\vertex}  \gg \sigma'_\mr{cl,\bubble}$ by orders of magnitude, 
hence the bubble contribution is irrelevant and $\sigma'_\mr{cl}$ is entirely dominated by the vertex contribution.
Moreover, both $T \sigma'_{\mr{cl}}$ and $T \sigma'_{\mr{cl,\vertex}}$
exhibit $\omega/T$ scaling [Figs.~\ref{fig:res}(b,d)],
the latter even cleaner than the former.
By contrast, $T\sigma'_{\mr{cl,\bubble}}$ does not exhibit scaling (not shown).

Large vertex contributions with the same scaling form as above have previously 
been reported \cite{Gleis2025} in a strange metal regime in the vicinity of a heavy-fermion QCP, found \cite{Gleis2024_PRX} in a cellular DMFT treatment of the 3-dimen\-sional periodic Anderson model. 
There, the uniform lattice conductivity 
$\sigma'_{\mr{latt}}(\omega,T)$ was approximated \cite{Gleis2025} by 
$\sigma'_{\mr{latt}} \simeq  \sigma'_{\mr{latt,\bubble}}
+\sigma'_{\mr{cl,\vertex}}$, involving its bubble contribution and a vertex contribution restricted to the cluster, neglecting longer-ranged terms. 
In the strange metal regime this yielded $\sigma'_{\mr{cl,\vertex}} \gg \sigma'_{\mr{latt,\bubble}}$, 
i.e.\ the lattice bubble contribution is irrelevant, $\sigma'_{\mr{latt}} \simeq \sigma'_{\mr{cl,\vertex}}$, and $\sigma'_\mr{latt} (0,T) \sim 1/T$.
This is in strong contrast to
MFL phenomenology, which assumes $\sigma'_{\mr{latt}} \simeq \sigma'_{\mr{B,latt}}$.
The periodic Anderson model was thus found to yield an intrinsic, short-ranged current decay mechanism without invoking disorder.

In the present case, very deep in the NFL region, for $\omega, T \lesssim 10^{-2}/\TNFL$, 
we likewise find $\sigma'_{\mr{cl,\vertex}} \gg \sigma'_{\mr{latt,\bubble}}$. 
That suggests a similar conclusion, in particular a strange-metal conductivity $\sigma'_{\mr{latt}}(0,T) \propto 1/T$.
However, we are unable to establish this with certainty,
since our results for the scaling behavior of the self-energy
$\Sigma(\omega,T)$, which determines that of $\sigma'_{\mr{latt,\bubble}}(\omega,T)$,
are not sufficiently accurate for very small $\omega$ and $T$ \cite{limitedaccuracy}.

For somewhat higher temperatures and frequencies, 
$\omega, T \in [10^{-2}\TNFL, \TNFL]$, we find evidence for a MFL self-energy at the node, discussed in detail below, leading to $\sigma'_{\mr{latt,\bubble}}(0,T) \propto 1/T$ behavior, as discussed with Fig.~\ref{fig:Bubble} of the End Matter below.
We expect that a more accurate determination of $\Sigma(\omega,T)$ \cite{limitedaccuracy} will confirm MFL behavior at least to leading order in the scattering 
of electrons off the spin fluctuations discussed in Fig.~\ref{fig:Sz_omega}; then, \ $\sigma'_{\mr{latt,\bubble}}(0,T)$ will
not be more singular than $1/T$. All in all, our computations thus strongly suggest a strange-metal conductivity with $\sigma'_{\mr{latt}}(0,T) \propto 1/T$.

\begin{figure}
    \centering
    \includegraphics[width=\linewidth]{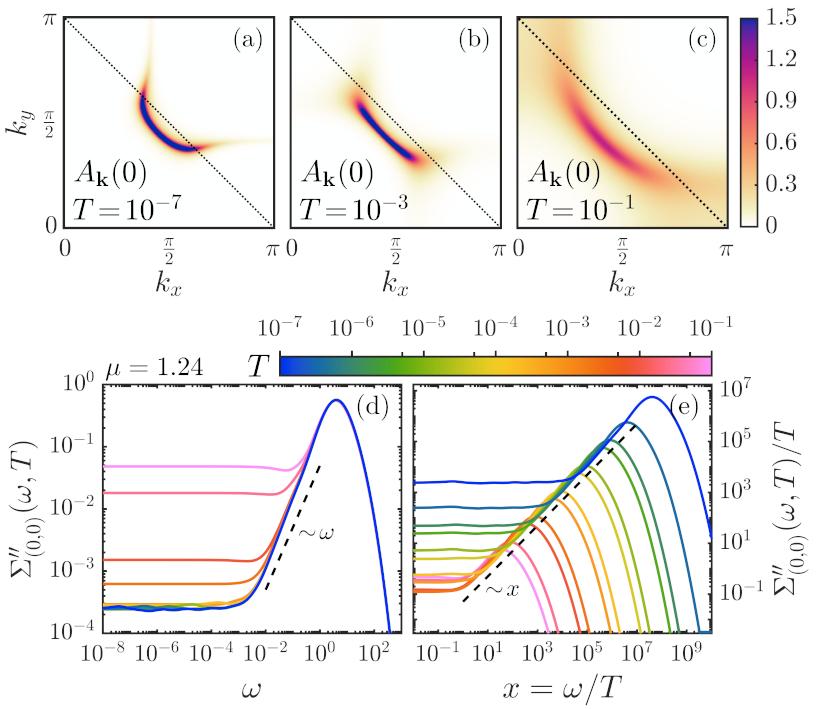}
    \caption{The  spectral function $A_{\bk}(\omega=0)$,  computed at $\mu=1.24$ for (a) the PG phase at $T=10^{-7}t$, (b) the NFL at $T=10^{-3}t$ and (c) the NFL at $T=10^{-1}t$. The black dotted line indicates the antiferromagnetic zone boundary.
    (d) The spectral part of the self-energy at the nodal patch $\bK=(0,0)$, $\Sigma''_{(0,0)}(\omega,T)$ vs.\ $\omega$, and (e) $\Sigma''_{(0,0)}(\omega,T)/T$ vs.\ $x=\omega/T$. 
    } \label{fig:Ak0_Linterp-revised} \vspace{-5mm}
\end{figure}

\textit{Electronic spectra and self-energy.---}
Experimentally, it is established that the cuprates exhibit a Fermi surface with at least somewhat coherent electronic excitations in the strange metal region.
Around optimal doping and in the underdoped region, these are mostly present in the nodal region, while the antinodal region appears to be much more incoherent, leading to Fermi arcs~\cite{Olson1990,Valla1999,Kanigel2006,Kohsaka2008,He2014,Kondo2015,Drozdov2018}.
Our DCA+NRG computations reproduce this behavior, 
see Figs.~\ref{fig:Ak0_Linterp-revised}(a-c), which display momentum-dependent, zero-frequency spectral functions $A_{\bk}(0)$ at different temperatures.
(We obtained their smooth momentum dependence using Liouvillian
interpolation~\cite{Pelz2026_Linterp} of the raw, momentum-discontinuous DCA self-energy (cf.\ Fig.~\ref{fig:Ak0_cluster} of the supplement~\cite{supplement}).
The Fermi arcs of Figs.~\ref{fig:Ak0_Linterp-revised}(a-c) show
the typical behavior observed experimentally ~\cite{Kanigel2006}, becoming longer and more coherent as the temperature is lowered.
Our choice of $\mu=1.24$ yields a PG metal at the lowest temperatures. As established in Ref.~\cite{Pelz2026_QCP}, this PG metal exhibits \FermiL\ quasiparticles with a reconstructed Fermi surface and a quasiparticle composition and weight with a strong notal-antinodal dichotomy.

The degree of coherence in the nodal region is determined by the frequency and temperature dependence of the imaginary part of the self-energy, $\mr{Im}\Sigma_{\bk}(\omega,T)$.
In our DCA approach, the self-energy at the node is determined by the self-energy of the $\bK = (0,0)$ patch, $\Sigma_{(0,0)}(\omega,T)$. For $T \gtrsim 10^{-3}$, its spectral part scales as $\Sigma'' \sim \omega$
for $\omega \gg T$, see Fig.~\ref{fig:Ak0_Linterp-revised}(d),
and roughly as $\Sigma'' \sim T$ for $\omega < T$, see Fig.~\ref{fig:Ak0_Linterp-revised}(e), where $\Sigma''/T$ vs.\ $x = \omega/T$ depends only weakly on 
$T$ for  $x< 1$ and $T > 10^{-3}$. (For  $x < 1 $ and $T< 10^{-3}$, however, 
the dependence of $\Sigma''$ on $T$ saturates
[Fig.~\ref{fig:Ak0_Linterp-revised}(d)] 
and $\Sigma''/T$ vs.\ $x$ scaling breaks down [Fig.~\ref{fig:Ak0_Linterp-revised}(e)]; this likely reflects numerical inaccuracies in the nodal self-energy 
at very low frequencies \cite{limitedaccuracy}.)
These results suggest that the nodal self-energy shows
MFL behavior, $\Sigma''_{(0,0)} (\omega,T) \sim {\rm{max}}(\omega,T)$, but more accurate future computations are needed to solidify this claim.

Our results are consistent with Ref.~\cite{Wu2022_PNAS}, where a linear-in-$T$ electronic scattering rate was found using DCA with 8 patches together with a Quantum Monte Carlo~(QMC) impurity solver, in a similar parameter regime for $T > 0.02$.
These 8-patch results were found to be consistent with 4-patch and 16-patch calculations, i.e.\ independent of cluster size.
Experimentally, $\mr{Im}\Sigma_{\bk}(\omega,T)$ has been determined from momentum or energy distribution curves, and was found to be linear in frequency and temperature at the node and at optimal doping~\cite{Olson1990,Valla1999,Chang2008,Reber2019}, consistent with MFL phenomenology and with DCA predictions for the Hubbard model.

\textit{Conclusion and outlook.---}
Using four-patch DCA+ NRG, we studied scaling in the quantum critical region of the pseudogap quantum critical point of the Hubbard model.
We find dynamical $\omega/T$ scaling for local spin and cluster-current susceptibilities, strong vertex contributions to the cluster conductivity, and evidence for a MFL self-energy in the nodal region.
Our results strongly support a strange metal optical response in the quantum critical region. Important future goals are a more accurate computation of the DCA+NRG self-energy; a more in-depth analysis of the scaling of the optical response akin to Ref.~\cite{Michon2019}; and computing the NFL behavior of other observables such as the specific heat, the Hall response, or thermoelectric transport coefficients.

\begin{acknowledgments}
We thank Antoine Georges, Seung-Sup Lee, Sayantan Roy, and Alexei Tsvelik for valuable discussions. We acknowledge the Gauss Centre for Supercomputing e.V. (www.gauss-centre.eu) for funding this project by providing computing time on the GCS Supercomputer SUPERMUC-NG at Leibniz Supercomputing Centre (www.lrz.de). We also acknowledge additional computational resources provided by the Arnold Sommerfeld Center for theoretical physics (www.theorie.physik.uni-muenchen.de). This work was supported in part by the Deutsche Forschungsgemeinschaft under grants INST 86/1885-1 FUGG, LE 3883/2-2, DE 730/16-2, and Germany’s Excellence Strategy EXC-2111 (Project No. 390814868). It is part of the Munich Quantum Valley, supported by the Bavarian state government with funds from the Hightech Agenda Bayern Plus. The National Science Foundation supported JvD in part under PHY-1748958, and GK under Grant No. DMR-1733071.
AG acknowledges support from the Abrahams Postdoctoral Fellowship of the Center for Materials Theory at Rutgers University.

\end{acknowledgments}
\bibliography{bibfile} 
\clearpage

\setcounter{secnumdepth}{2} 

\bigskip
\bigskip
\onecolumngrid
\begin{center}
\textbf{\large End Matter}
\end{center}
\bigskip
\vspace{-0.2cm}
\twocolumngrid

\section{Imaginary part of the optical conductivity}\label{app:ImSigma}

\begin{figure}[t]
    \centering
    \includegraphics[width=\linewidth]{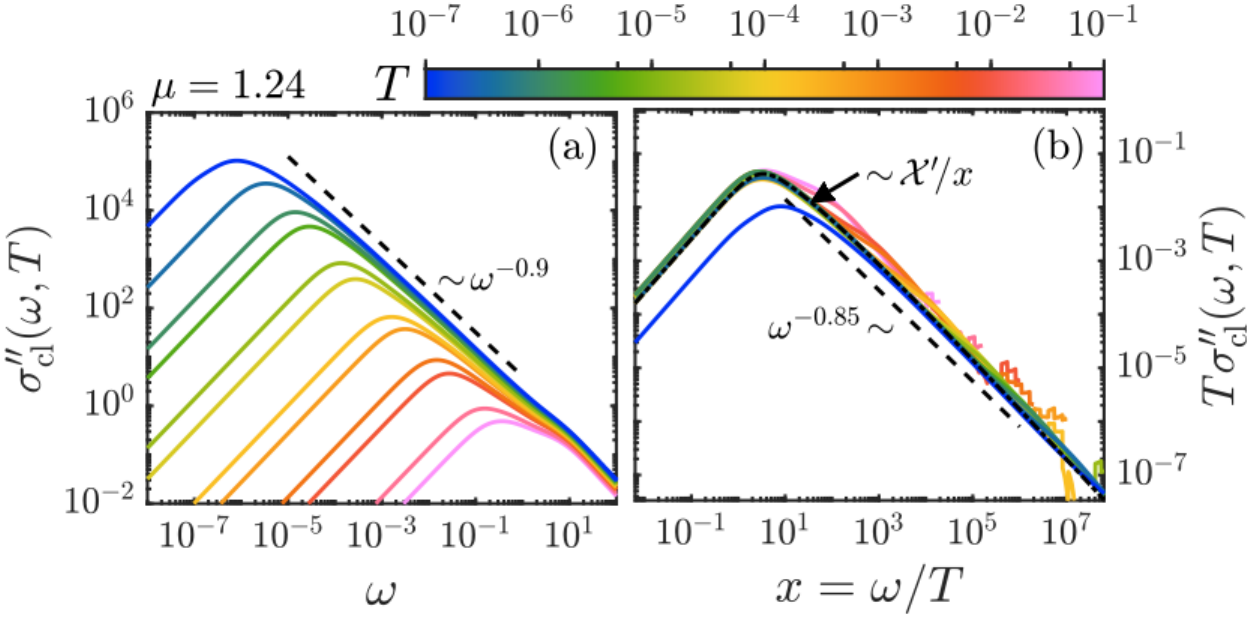}
    \caption{Spectral part of the cluster optical conductivity, $\sigma''_\mr{cl}(\omega,T)$, at $\mu=1.24$. (a) $\sigma''_\mr{cl}$ vs.\ $\omega$ for 12 different temperatures, and (b) $T\sigma''_\mr{cl}$ vs.\ $x=\omega/T$, 
    showing a scaling collapse onto $\mathcal{X}''(x)/x$ (black dash-dotted curve). Black dashed lines indicate power law behavior.
    }
    \label{fig:ImSigma}
\end{figure}

In the main text, we discussed $\sigma'_\mr{cl}(\omega,T)$, the real part of the cluster optical conductivity. Here, we discuss the corresponding spectral part, $\sigma''_\mr{cl}(\omega,T)$, obtained via Kramers-Kronig. Figure~\ref{fig:ImSigma}(a) shows $\sigma''_\mr{cl}$ vs.\ $\omega$; for $\omega> T$, it decays slightly more slowly than $\omega^{-0.9}$ (dashed line).
Figure~\ref{fig:ImSigma}(b) shows the corresponding scaling behavior of $T \sigma''_\mr{cl}$ vs.\ $x=\omega/T$ (with some numerical artifacts at very 
high frequencies stemming from the Kramer-Kronig computation).
Interestingly, the scaling collapse curve shows rather pure  power-law decay,  $\sim x^{-0.85}$ (dashed line).
Similar exponents have been observed experimentally in Bi2212 and YBCO \cite{Azrak1994,Baraduc1996}. Such a decay exponent slight smaller than 1 can be indicative of a logarithmic renormalization of the effective mass due to strong local vertex contributions, leading to logarithmic temperature scaling of the optical mass \cite{Michon2023}.

\section{Bubble contribution}

The bubble contribution to the optical conductivity is obtained from the $\mathbf{q} \rightarrow \mathbf{0}$ limit of the current–current correlator without vertex contributions. In this approximation, the conductivity is given by a sum over velocity-squared weights times products of Green’s functions (or spectral functions), describing independent particle–hole excitations driven by a uniform electric field, with interactions entering only through the dressed propagators.

In this work, we considered two approaches for approximately computing
the bubble contributions, using Green's functions computed on the cluster or
the lattice, resulting in $\sigma'_{\mr{cl}, \bubble}(\omega,T)$ or $\sigma'_{\mr{latt},\bubble}(\omega,T)$, respectively. The vertex contribution to the cluster  conductivity, shown in the main text, is obtained as 
$\sigma'_{\mr{cl},\vertex} = \sigma'_\mr{cl} - \sigma'_{\mr{cl},\bubble}$. For computational details, see the Supplemental Material~\cite{supplement}.

\Fig{fig:Bubble} shows the two types of bubble contributions as functions of frequency, with colors indicating temperature. 
The cluster bubble contribution $\sigma'_{\mr{cl},\bubble}$ [\Fig{fig:Bubble}(a)] is negligible compared to the full cluster conductivity $\sigma'_{\mr{cl}}$ [cf.\ \Fig{fig:res}(a)], particularly at low temperatures, indicating that $\sigma'_{\mr{cl}}$ is dominated by vertex contributions [cf.\ \Fig{fig:res}(c)]. The lattice bubble 
$\sigma'_{\mr{latt},\bubble}$ [\Fig{fig:Bubble}(c)] is several orders of magnitude larger than the cluster bubble $\sigma'_{\mr{cl},\bubble}$, but remains small relative to the full cluster conductivity $\sigma'_{\mr{cl}}$ [\Fig{fig:Bubble}(a)]. At the lowest temperature, $\sigma'_{\mr{latt},\bubble}$ exhibits the emergence of a precursor of a Drude peak, consistent with the onset of coherent transport in the pseudogap regime outside the quantum critical cone.
\begin{figure}[t]
    \centering
    \includegraphics[width=\linewidth]{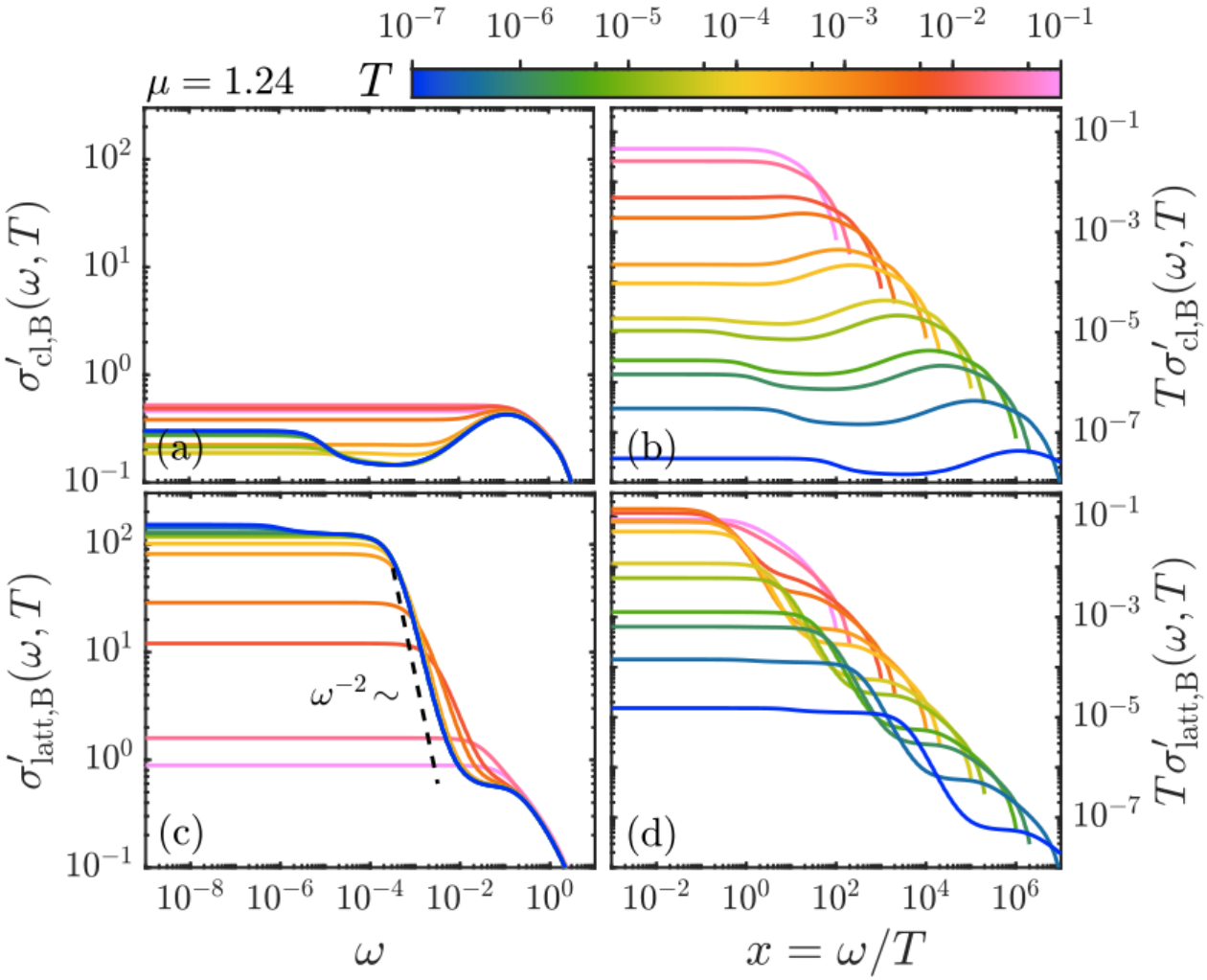}
    \caption{Two approximations for the bubble contribution to the
    real part of the optical conductivity: (a,b) $\sigma'_{\mr{cl},\bubble}(\omega,T)$, computed on the cluster, and  (c,d) $\sigma'_{\mr{latt},\bubble}(\omega,T)$, computed for the lattice. (a,c) show $\sigma'_\bubble$ vs.\ $\omega$ for several temperatures, (b,d) show $T \sigma'_\bubble$ vs. $x = \omega/T$. For ease of comparison, panels (a,c) use the same range for the ordinate, likewise for (b,d). 
    }
    \label{fig:Bubble}
\end{figure} 

\section{Current-Current contributions}

\begin{figure}[t]
    \centering
    \includegraphics[width=\linewidth]{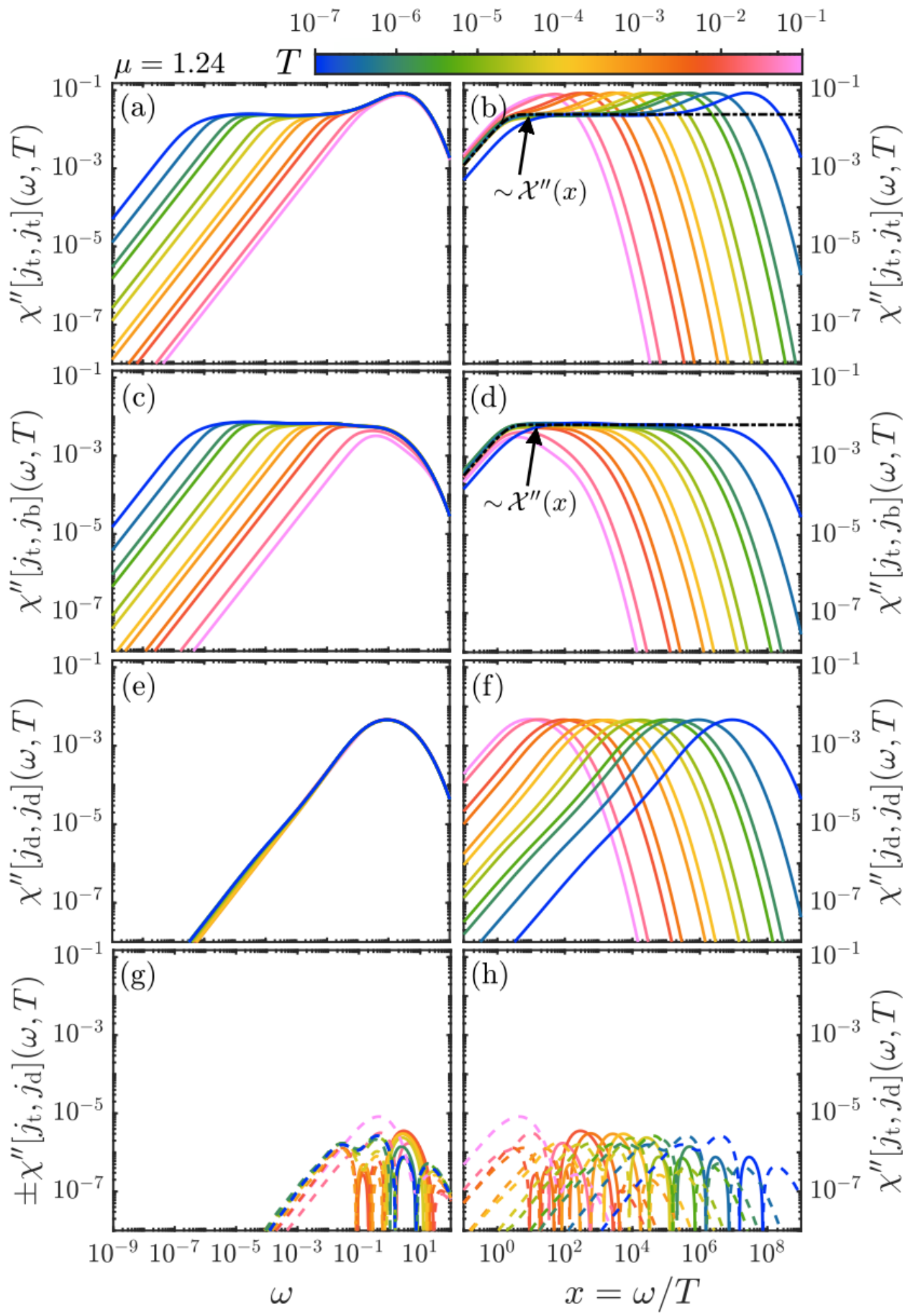}
    \caption{Four types of contributions to the spectral part, $\chi''[j^x_{\mr{cl}}](\omega,T)$, of the cluster-current correlator, shown as functions of $\omega$ for 12 temperatures on the left, and as functions of $x=\omega/T$ on the right: 
    (a,b) same-rung, (c,d) opposite-rung, (e,f) 
    all-diagonal, and (g,h) rung-diagonal correlators.
    In (g,h) sign changes in $\chi''[j_\mr{t},j_\mr{d}]$ are accounted for using solid or dashed lines to show $+\chi''$ or $-\chi''$, respectively. For ease of comparison, all panels use the same range for the ordinate. Black dashed lines indicate linear frequency behavior, while black dash-dotted lines correspond to the scaling function $\mc{X''}(x)=\mc{X}''_0 \tanh(x/2)$, with $\mc{X}''_0\simeq0.024$ for $\chi''[j_\mr{t},j_\mr{t}]$ and $\mc{X}''_0\simeq0.0067$ for $\chi[j_\mr{t},j_\mr{b}]$.
    }
    \label{fig:jcl_contributions}
\end{figure}
In this section, we discuss the dependence of the cluster current-current correlation function $\chi''[j^x_{\mr{cl}}](\omega,T)$  on its individual contributions. 
The definition \eqref{eqs:operators} contains three types of terms, 
$j_{\mr{cl}}^{x} = j_{\mr{t}} + j_{\mr{b}} + j_{\mr{d}}$,
describing hopping processes on the square cluster proceeding left-to-right  
along the top (t) rung, the bottom (b) rung, or along the two diagonals (d). 
Thus, the cluster-current susceptibility $\chi[j^x_\mr{cl}]$ contains
nine terms, which we group into four types: 
same-rung (\mbox{t-t}, \mbox{b-b}), opposite-rung (\mbox{b-t}, \mbox{t-b}), 
all-diagonal (\mbox{d-d}),  
and rung-diagonal (\mbox{t-d}, \mbox{b-d}, \mbox{d-t}, \mbox{d-b}) correlators, e.g.\
\begin{align}
    \chi[j_\mr{t},j_\mr{b}](\omega,T) = -\ii\int_0^\infty \dd t\, e^{\ii(\omega+\ii0^+)t} \left\langle [j_\mr{t} (t), j_\mr{b} ]\right\rangle . 
\label{eq:retarded-appendix}
\end{align}
Figure~\ref{fig:jcl_contributions} shows the $(\omega,T)$ dependence of 
one representative from each type (left panels), and its dependence on $x=\omega/T$
(right panels). The same-rung correlators [\Figs{fig:jcl_contributions}(a,b)], i.e.\ the most local ones, give the  dominant contribution to the plateau in $\chi''[j^x_\mr{cl}]$
underlying Planckian scaling. The opposite-rung correlators also give plateau contributions, albeit smaller by roughly a factor of 2 [\Figs{fig:jcl_contributions}(c,d)]. The all-diagonal contributions show no plateaus at all [\Figs{fig:jcl_contributions}(e,f)], while the rung-diagonal contributions, the only ones exhibiting sign changes, are several orders of magnitude smaller than all others
[\Figs{fig:jcl_contributions}(g,h)]. 

We conclude that the plateau underlying Planckian scaling is dominated by the most short-ranged current-current correlations on the cluster, highlighting the strongly local character of the underlying quantum critical dynamics. It thus stands to reason that a marked plateau and Planckian scaling will also be found when studying larger clusters---although these will involve additional current correlators of longer range, their contributions will likely be even more sub-dominant than those considered here.


\clearpage

\thispagestyle{empty}

\setcounter{equation}{0}
\setcounter{figure}{0}
\setcounter{page}{1}

\renewcommand{\theequation}{S\arabic{equation}}
\renewcommand{\thefigure}{S\arabic{figure}}
\renewcommand{\thepage}{S\arabic{page}}

\setcounter{secnumdepth}{2} 
\renewcommand{\thefigure}{S\arabic{figure}}
\setcounter{figure}{0}
\setcounter{section}{0}
\setcounter{equation}{0}
\renewcommand{\thesection}{S-\Roman{section}}
\renewcommand{\theequation}{S\arabic{equation}}
%

\title{Supplemental Material for 
``{Dynamical scaling near the pseudogap quantum critical point of the two-dimensional Hubbard model}''}

\date{\today}
\maketitle

\section{Scaling function}\label{app:planck}

In the main text, we discussed Planckian $\omega/T$ scaling 
for two correlators whose spectral parts exhibit a plateau: $\chi[S^z_\br]$ and $\chi[j^x_\mr{cl}]$, the correlators for the 
local spin and the cluster-current in the $x$-direction, defined in Eqs.~\eqref{eqs:operators}.

Our NRG impurity solver directly computes the corresponding real frequency spectra $\chi''[O](\omega,T)$. We use the discrete NRG data for the latter as input to evaluate the respective imaginary-time correlators $\langle O(\tau) O^\dagger\rangle$ via \Eq{eq:tau_transform}, thereby avoiding broadening artifacts. 

In the imaginary-time domain, a correlator of the form
\begin{align}
    \langle O(\tau) O^\dagger \rangle 
    \sim \frac{\pi T}{\sin(\pi T\tau)}
    \label{eq:chi_tau}
\end{align}
represents scale-invariant finite-temperature behavior, where temperature sets the only energy scale and the dependence enters through the dimensionless combination $\pi T\tau$. The form \eqref{eq:chi_tau} has the imaginary-time periodicity properties required for finite-temperature correlators and can be viewed as the thermal generalization of a zero-temperature $1/\tau$ correlator. 

The spectral function  associated with the correlator \eqref{eq:chi_tau} has the form $\chi''[O](\omega,T) = \mc{X}''(x)$, where 
\begin{align}
    \mc{X}''(x) = \mc{X}''_0\,\tanh (x/2) \, , 
    \quad x = \omega/T \, . 
    \label{eq:chi_fit}
\end{align}
This has properties characteristic of Planckian scaling: linear in $x$ for $x\ll1$ and  saturation to a plateau at $x \gg 1$. The parameter $\mc{X}''_0$ sets the overall scale. 

We define the real part of the correlator, $\chi'[O](\omega,T)$, by the Kramer-Kronig relation
\begin{align}
    \mc{X}'(x,y) &= \mc{P}\int_{-y}^y \mr{d}\widetilde{x} \frac{\mc{X}''(\widetilde{x})}{x-\widetilde{x}} ,
    \label{eq:ReX}
\end{align}
introducing $y$ as an ultraviolet cutoff 
to regularize the integral, which diverges as $-\ln y$ for
$y\to \infty$. To isolate the divergent part  \cite{Gleis2025}, we use 
$\frac{1}{x-\widetilde x} = - \frac{1}{\widetilde x} + \frac{x}{(x-\widetilde x) 
\widetilde x} $ and write 
$\mc{X}'(x,y) \simeq  \mc{X}_0'(y) + \mc{X}'(x)$, with 
\begin{subequations}
\begin{align}
\label{eq:regularizeX'-b}
    \mc{X}_0'(y) & = 
    -\mc{P}\int_{-y}^y\mr{d}\widetilde{x} \frac{\mc{X}''(\widetilde{x})}{\widetilde{x}} \, , 
    \\ 
    \label{eq:regularizeX'-a}
    \mc{X}'(x) &= \phantom{-} \mc{P}\int_{-\infty}^{\infty}\mr{d}\widetilde{x} \frac{x\mc{X}''(\widetilde{x})}{(x-\widetilde{x})\widetilde{x}} \, . 
\end{align}
\end{subequations}
Here, $\mc{X}_0'(y)
\simeq \mc{X}'(0,y)$ represents the singular, ``static'' contribution, and we took $y\to \infty$ in 
\Eq{eq:regularizeX'-a}. All in all, we thus obtain a scaling function of the form 
\begin{align}
    \mc{X}(x, y) 
    \simeq \mc{X}'_0 (y) 
    + \mc{X}' (x) 
    -\mr{i}\pi\mc{X}'' (x) 
    \ .
    \label{eq:X_tanh}
\end{align}
In the main text, we employed this form 
to fit correlators $\chi(\omega,T)$
computed with DCA+NRG, using $y= \TNFL/T$
as ultraviolet cutoff \cite{Gleis2025}. Then, both 
$\chi''(\omega,T) = \mc{X}''(x)$
and $\chi'[O](\omega,T) - \chi'(0,T) = \mc{X}'(x)$
show scaling with $x=\omega/T$, 
and $\chi'[O](0,T)= \mc{X}'(\TNFL/T)$
describes the static part. When applied to the correlators $\mc{X}[S_\bri ^z]$
and  $\mc{X}[j_{\mr{cl}}^x]$ discussed in the main text,
fits of our numerical data for $\chi''(\omega,T)$ to \Eq{eq:chi_fit}, with  $\mc{X}''_0$ as fit parameter, yield $\mc{X}''_0[S_\bri ^z]\simeq 0.132$ and $\mc{X}''_0[j_{\mr{cl}}^x]\simeq 0.0189$.

\section{Spectral functions}\label{app:SpecFunc}

\begin{figure*}[t]
    \centering
    \includegraphics[width=\linewidth]{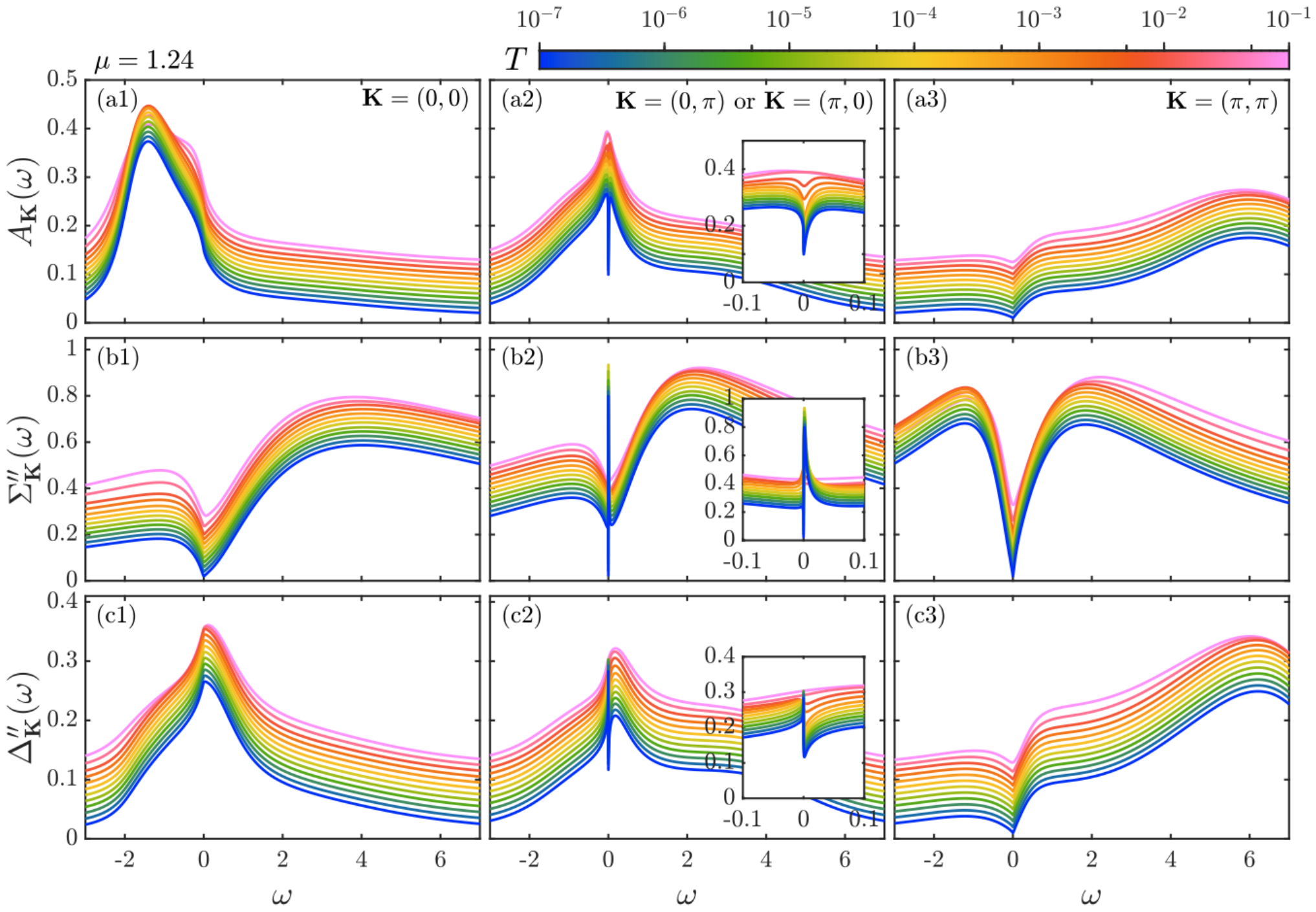}
    \caption{Spectral functions of (a) the Green's function, (b) the self-energy, and (c) the hybridization at $\mu=1.24$ as a function of $\omega$ for several temperatures $T$, denoted by different colors. Columns (1-3) depict DCA patch momenta $\bK=(0,0)$, $\bK=(0,\pi)$ or $\bK=(\pi,0)$, and $\bK=(\pi,\pi)$, respectively. For visual clarity, spectral functions for successive temperatures are vertically offset by 0.01 for $A_\bK$ and $\Delta''_\bK$, and by 0.02 for $\Sigma''_\bK$. The insets for the anti nodal points show a zoom around $\omega=0$.
    }
    \label{fig:SpectralFunctions}
\end{figure*}
\begin{figure}[t]
    \centering
    \includegraphics[width=\linewidth]{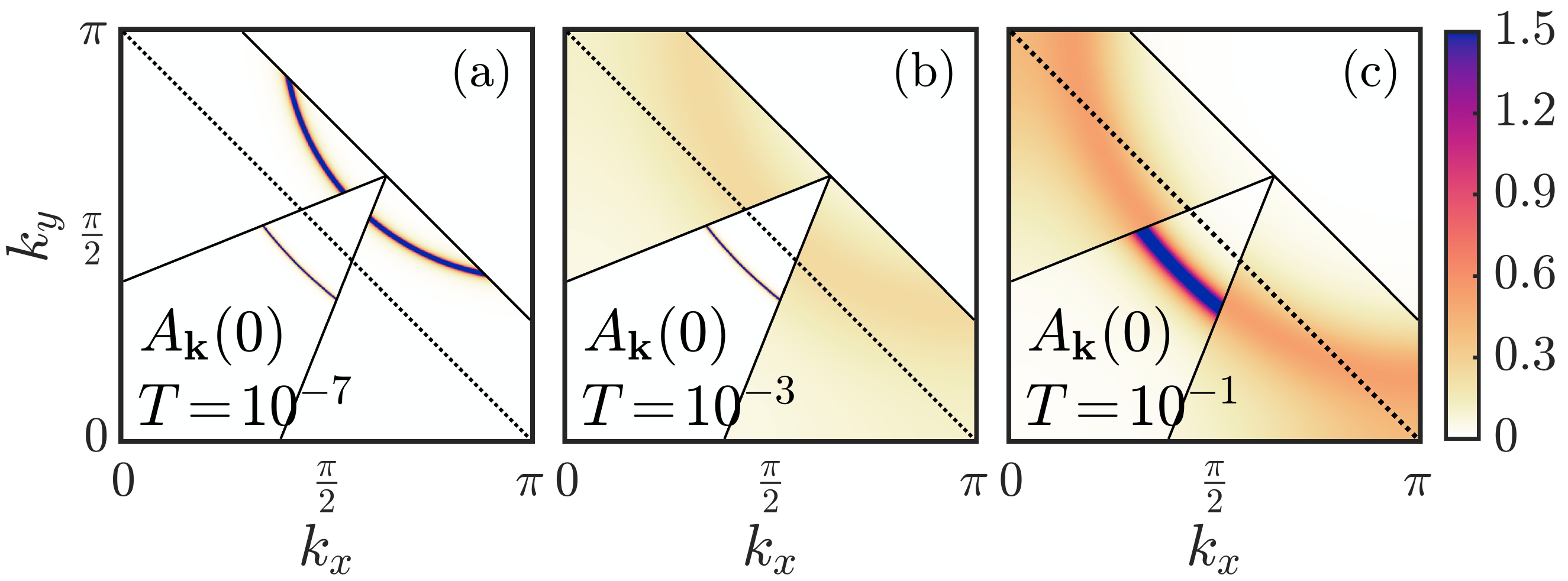}
    \caption{Analogous to \Fig{fig:Ak0_Linterp-revised}, but now showing spectral functions $A_\bk(\omega=0)$ computed using the non-interpolated DCA results for the patch self-energies $\Sigma_\bK(\omega = 0)$. Black lines mark DCA patch boundaries.}
    \label{fig:Ak0_cluster}
\end{figure}

In this section, we analyze the temperature evolution of the spectral parts of the patch Green’s functions $G_{\bK}$, self-energies $\Sigma_{\bK}$, and hybridization functions $\Delta_{\bK}$ obtained from our DCA calculations at chemical potential $\mu=1.24$. We denote their spectral parts, defined as $X_{\bK}''(\omega) =-\frac{\mr{i}}{2\pi}[X_{\bK}^{\pdag}(\omega)-X_{\bK}^{\dagger}(\omega)]$, by $A_{\bK}$, $\Sigma''_\bK$ and 
$\Delta''_\bK$, respectively. These are shown in the first, second and third rows of  \Fig{fig:SpectralFunctions}, with each column corresponding to a specified DCA patch  momentum $\bK$. 

In the nodal patches $\bK=(0,0)$ and $\bK=(\pi,\pi)$, which dominate the high- and low-frequency regimes and are associated with the upper and lower Hubbard bands, respectively, the spectral functions $A_{\bK}$, $\Sigma''_\bK$ and $\Delta_\bK$ all depend only very weakly on temperature. This implies that  coherence is preserved as a function of temperature along the nodal direction.

By contrast, for the spectral functions for the degenerate antinodal patches, $\bK=(0,\pi)$ and $\bK=(\pi,0)$, exhibit a pronounced temperature dependence. For $A_{(0,\pi)}$, we observe the well-known opening of a pseudogap as $T$ decreases [\Fig{fig:SpectralFunctions}(a2)]. Correspondingly, $\Sigma''_{(0,\pi)}$ develops a sharp peak, signaling the loss of coherence [\Fig{fig:SpectralFunctions}(b2)]. For  the hybridization function, a step-like feature emerges with decreasing temperature, leading to a strong particle–hole asymmetry at $\omega=0$ [\Fig{fig:SpectralFunctions}(c2)].

Figure~\ref{fig:Ak0_cluster} shows results for the spectral functions
$A_\bk(\omega=0)$ computed using the raw (non-interpolated) DCA results for 
the patch self-energies $\Sigma_\bK(\omega=0)$ at three temperatures. (Figure~\ref{fig:Ak0_Linterp-revised} was computed analogously, 
but using a self-energy $\Sigma_\bk$ obtained by Liouvillian interpolation \cite{Pelz2026_Linterp}.) Consistent with the spectral functions shown in \Figs{fig:SpectralFunctions} and \ref{fig:Ak0_Linterp-revised}, the nodal patches remain essentially unaffected by the change in temperature. In contrast, the antinodal patches become strongly broadened as the temperature increases.

\section{Computation of the bubble contributions}

In this section, we detail the formulas used to compute the bubble contributions to the optical conductivity, 
$\sigma'_{\mr{cl},\bubble}(\omega,T)$ and $\sigma'_{\mr{latt},\bubble}(\omega,T)$, computed for the cluster or the lattice, respectively.
For numerical accuracy, we employ the integration method for the convolution of the bubble contribution introduced in Ref.~\cite{Gleis2025}.

\subsection{Lattice Bubble}

A conclusive determination of the resistivity scaling requires evaluating the lattice bubble rather than the local cluster bubble. In particular, the local vertex contributions computed on the cluster must dominate the low-frequency behavior over the lattice bubble term to yield the same linear-in-$T$ scaling.

However, resolving this regime is numerically challenging. Solving a four-site cluster impurity within NRG is computationally demanding due to the exponential growth of the Hilbert space with cluster size. As a result, the very low-frequency regime ($\omega \lesssim T$) of the self energy, which is only visible on a logarithmic frequency grid, is not sufficiently reliable to accurately determine the lattice bubble contribution. Further numerical improvements are therefore required to establish scaling behavior beyond the local cluster level.

Nevertheless, we compute and show $\sigma_{\mr{latt},\bubble}(\omega,T)$ compared to the cluster bubble for future reference. The lattice contribution to the optical conductivity is computed within the bubble approximation by explicitly retaining the dispersion dependence within each momentum patch. The total conductivity is obtained as an average over cluster momenta,
\begin{align}
    \sigma_{\mr{latt,\bubble}}(\omega) = \frac{1}{N_{\mr{p}}}\sum_{\bK} \sigma^{\mr{latt,\bubble}}_{\bK}(\omega)\, ,
\end{align}
where each patch contribution $\sigma^{\mr{latt,\bubble}}_{
\bK}(\omega)$ is expressed as an integral over the band energy $\epsilon$,
\begin{align}
    \sigma^{\mr{latt,\bubble}}_{\bK}(\omega) = 2\pi e^2 \int d\epsilon \, \Phi^{xx}_{\bK}(\epsilon) \, \tilde{\sigma}^{\mr{latt,\bubble}}_{\bK}(\epsilon,\omega) \, , \label{eq:sigK_latt}
\end{align}
Here, $\Phi^{xx}_{\bK}(\epsilon)$ is the energy-resolved transport function, defined as the patch average of the squared band velocity weighted by the density of states,
\begin{align}
    \Phi^{xx}_{\bK}(\epsilon) = \int_{V_{\bK}} \frac{\mathrm{d}\bk}{V_{\bK}} \, \left(v^x_{\bk}\right)^2 \delta(\epsilon-\epsilon_{\bk}) \, ,
\end{align}
where $v^x_{\bk}= \partial \epsilon_{\bk}/\partial k_x$ is the band velocity in the $x$-direction, and the integral is over momentum patch $\bK$ with patch volume $V_\bK$. The transport function $\Phi^{xx}_{\bK}(\epsilon)$ captures the geometric contribution of the electronic dispersion to the optical response and weights the spectral functions involved in 
$ \tilde{\sigma}^{\mr{latt,\bubble}}_{\bK}$ [see \Eq{eq:sigma-lattice-bubble}] according to their current-carrying capability within each patch.

The dynamical part of the conductivity is encoded in
the particle–hole bubble at band energy $\epsilon$, 
\begin{align}
\label{eq:sigma-lattice-bubble}
    \tilde{\sigma}^{\mr{latt,\bubble}}_{\bK}(\epsilon,\omega) =\!\! \int \!\mathrm{d}\widetilde{\omega}\, \frac{f(\widetilde{\omega})\!-\!f(\widetilde{\omega}\!+\!\omega)}{\omega} A_{\bK}(\epsilon,\widetilde{\omega}) A_{\bK}(\epsilon,\widetilde{\omega}\!+\!\omega) \, .
\end{align}
Here, $f(\omega)$ is the Fermi–Dirac distribution, and the patch spectral function
\begin{align}
    A_{\bK}(\epsilon,\omega) = -\frac{1}{\pi} \mr{Im}\bigg[ \frac{1}{\omega + \mu - \epsilon - \Sigma_{\bK}(\omega)}\bigg]
\end{align}
depends on the patch self-energy $\Sigma_{\bK}(\omega)$, which momentum-independent within each patch. Finally, the dc conductivity is obtained from the zero-frequency limit,
\begin{align}
    \sigma^{xx} = \lim_{\omega \to 0} \sigma(\omega)\, .
\end{align}

The above formulation effectively separates the geometric (band-structure) contribution, captured by $\Phi^{xx}_{\bK}(\epsilon)$, from the many-body dynamics encoded in the spectral functions. Compared to the cluster bubble discussed below, this approach replaces the patch-averaged velocity factor by an energy-resolved transport function, thereby improving numerical accuracy and capturing intra-patch dispersion effects.

\subsection{Cluster Bubble}
If the cluster contribution to the local optical conductivity is computed within the bubble approximation, one uses formulas analogous to but simpler than those above: 
\begin{align}
    \sigma_{\mr{cl,\bubble}}(\omega) & =\frac{1}{N_{\mr{p}}}\sum_{\bK}\sigma^{\mr{cl,\bubble}}_{\bK}(\omega) \, ,
\\ 
    \sigma^{\mr{cl,\bubble}}_{\bK}(\omega) & =2\pi e^2 \langle\Phi^{xx}_{\bK}\rangle \tilde{\sigma}^{\mr{cl,\bubble}}_{\bK}(\omega) \ ,
\\ 
    \tilde{\sigma}^{\mr{cl,\bubble}}_{\bK}(\omega) & =\!\! \int \!\mathrm{d}\widetilde{\omega}\, \frac{f(\widetilde{\omega})\!-\!f(\widetilde{\omega}\!+\!\omega)}{\omega} A_{\bK}(\widetilde{\omega}) A_{\bK}(\widetilde{\omega}\!+\!\omega) \, ,
    \label{eq:bubble_cluster}
\\     
    A_{\bK}(\omega) & = -\frac{1}{\pi} \mr{Im}\bigg[ \frac{1}{\omega + \mu - \epsilon_{\bK}-\Delta_{\bK}(\omega) - \Sigma_{\bK}(\omega)}\bigg] \, , 
\\ 
    \langle\Phi^{xx}_{\bK}\rangle & = \int_{V_{\bK}} \frac{\mathrm{d}\bk}{V_{\bK}}(v^x_\bK)^2  , 
    \qquad 
    \epsilon_\bK = \int_{V_{\bK}} \frac{\mathrm{d}\bk}{V_{\bK}} \epsilon_\bk \, . 
\end{align}
$\tilde{\sigma}^{\mr{cl,\bubble}}_{\bK}(\omega)$ is the particle–hole bubble, describing current–current correlations without vertex corrections; its the frequency convolution encodes the creation of particle–hole excitations at energy $\omega$, and the spectral functions occurring therein contain
$\epsilon_\bK + \Delta_\bK(\omega)$ in the denominator rather than $\epsilon_\bk$.

\clearpage

\end{document}